\documentclass{sigchi}

\toappear{\scriptsize Permission to make digital or hard copies of all or part of this work for personal or classroom use is granted without fee provided that copies are not made or distributed for profit or commercial advantage and that copies bear this notice and the full citation on the first page. Copyrights for components of this work owned by others than ACM must be honored. Abstracting with credit is permitted. To copy otherwise, or republish, to post on servers or to redistribute to lists, requires prior specific permission and/or a fee. Request permissions from permissions@acm.org. \\
{\emph{CHI '20, April 25--30, 2020, Honolulu, HI, USA.} } \\
Copyright is held by the owner/author(s). Publication rights licensed to ACM. \\
ACM ISBN 978-1-4503-6708-0/20/04\ ...\$15.00.\\
http://dx.doi.org/10.1145/3313831.3376264}

\usepackage[pdflang={en-US},pdftex]{hyperref}
\usepackage{balance}
\usepackage{graphics}
\usepackage[T1]{fontenc}
\usepackage{txfonts}
\usepackage{mathptmx}
\usepackage{color}
\usepackage{booktabs}
\usepackage{textcomp}
\usepackage{venndiagram}
\usepackage{microtype}
\usepackage{changepage}
\usepackage{wasysym}
\usepackage{enumitem}
\usepackage{footnote}
\usepackage[subtle]{savetrees}
\newcommand*\rot{\rotatebox{90}}
\usepackage{tabularx}

\usepackage{tikz}
\usetikzlibrary{shapes.geometric, arrows, positioning, snakes}
\tikzstyle{startstop} = [rectangle, rounded corners, minimum width=1.5cm, minimum height=0.5cm,text centered, draw=black, fill=red!30]
\tikzstyle{io} = [trapezium, trapezium left angle=70, trapezium right angle=110, minimum width=1.5cm, minimum height=0.5cm, text centered, draw=black, fill=blue!30]
\tikzstyle{process} = [rectangle, minimum width=1.5cm, minimum height=0.5cm, text centered, draw=black, fill=orange!30]

\def\plaintitle{Informing the Design of Privacy-Empowering Tools for the Connected Home}
\def\plainauthor{William Seymour, Martin J Kraemer, Reuben Binns, Max Van Kleek}
\def\plainkeywords{Technology Probe, Privacy-Empowering Technology, Network Disaggregator}

\makeatletter
\def\url@leostyle{%
  \@ifundefined{selectfont}{
    \def\UrlFont{\sf}
  }{
    \def\UrlFont{\small\bf\ttfamily}
  }}
\makeatother
\def\pprw{8.5in}
\def\pprh{11in}

\definecolor{linkColor}{RGB}{6,125,233}
\hypersetup{
  pdftitle={\plaintitle},
  pdfauthor={\plainauthor},
  pdfkeywords={\plainkeywords},
  pdfdisplaydoctitle=true,
  bookmarksnumbered,
  pdfstartview={FitH},
  colorlinks,
  citecolor=black,
  filecolor=black,
  linkcolor=black,
  urlcolor=linkColor,
  breaklinks=true,
  hypertexnames=false
}

\begin{document}

\title{Informing the Design of Privacy-Empowering Tools\\for the Connected Home}

\numberofauthors{4}
\author{%
  \alignauthor{William Seymour\\
    \affaddr{University of Oxford}\\
    \affaddr{Oxford, UK}\\
    \email{william.seymour@cs.ox.ac.uk}}\\
  \alignauthor{Martin J Kraemer\\
    \affaddr{University of Oxford}\\
    \affaddr{Oxford, UK}\\
    \email{martin.kraemer@cs.ox.ac.uk}}\\
  \alignauthor{Reuben Binns\\
    \affaddr{University of Oxford}\\
    \affaddr{Oxford, UK}\\
    \email{reuben.binns@cs.ox.ac.uk}}\\
  \alignauthor{Max Van Kleek\\
    \affaddr{University of Oxford}\\
    \affaddr{Oxford, UK}\\
    \email{max.van.kleek@cs.ox.ac.uk}}\\
}

\maketitle

\begin{abstract}
Connected devices in the home represent a potentially grave new privacy threat due to their unfettered access to the most personal spaces in people's lives. Prior work has shown that despite concerns about such devices, people often lack sufficient awareness, understanding, or means of taking effective action. To explore the potential for new tools that support such needs directly we developed Aretha, a privacy assistant technology probe that combines a network disaggregator, personal tutor, and firewall, to empower end-users with both the knowledge and mechanisms to control disclosures from their homes. We deployed Aretha in three households over six weeks, with the aim of understanding how this combination of capabilities might enable users to gain awareness of data disclosures by their devices, form educated privacy preferences, and to block unwanted data flows. The probe, with its novel affordances---and its limitations---prompted users to co-adapt, finding new control mechanisms and suggesting new approaches to address the challenge of regaining privacy in the connected home.
\end{abstract}




\section{Introduction}
\emph{My home is my castle}. This commonly-heard saying captures the importance and significance of the home to people's lives. The home is seen as an essential dominion over which one has complete control; a place providing essential safety and isolation from the outside world~\cite{marshall1972privacy}. Since early human civilisations, the home has played a role as a vital context for nearly every essential activity outside work~\cite{dickens1989human, parke1979children}.  The protection afforded by the home establishes a safe space to be genuinely one's self, permitting the `free unfolding of their personality'~\cite{eberle1997human}, a retreat where one can be vulnerable without fear of judgement, reproach, or attack by outsiders, and, finally, a place to seek shelter when recovering and resting.

The introduction of connected digital devices into the home has seen both positive and negative effects. Positive effects, such as increased safety and convenience, have been brought about by a variety of new capabilities, including advanced sensing and delegation, as exemplified by the likes of `smart' security systems and virtual assistants. But these very capabilities are also engendering a second set of effects, comprising concerns that such devices are threatening the core values of safety, privacy, and isolation of the home by introducing new vectors through which external entities can surveil the people and activities therein. These `foot in the door' devices are seen as bringing the kinds of surveillance-capitalism derived mass data collection, profiling, and targeted advertising that have thus far pervaded other online spaces into the most private physical spaces people need to thrive~\cite{pierce2019smart, zuboff2015big}.

The ultimate consequences of intrusion by connected devices into the private spaces of the home has yet to be fully understood. Researchers and journalists have documented the ways that devices are seen as creepy~\cite{fruchter2018consumer}, or make users feel constantly monitored or spied upon~\cite{spied,spied2}. Recent scandals involving products revealed to have hidden sensors not disclosed to end-users, including microphones in smart mattresses~\cite{smartmattress} and home hubs~\cite{microphones}, have further heightened anxieties around digital devices in the home.  Such incidents have caused end-users, consumer rights advocates, and privacy and security researchers to press for restrictions on data collection, and the establishment of mandatory software quality, security standards, and duties of care~\cite{javid2019online}.

Addressing privacy in the home is challenging for at least three interrelated reasons. End-users generally lack an awareness of how their devices collect data about them, including the kinds of data collected, and the ways such data are disclosed and used by various first- and third-party entities~\cite{gerber2018foxit,vankleek2017better}. Second, even when such data uses are made apparent, individuals lack both breadth and depth of knowledge necessary to form informed preferences about such disclosures.  Such knowledge includes both a broad, contextual understanding of data sharing norms, uses, and data protection requirements, on one hand, and specific, detailed knowledge on the other, such as of disclosure risks, or the reputations, business models, and security practices of the companies handling their data~\cite{Naeini2017,Zheng2018a}. Finally, even when such preferences have been formed, users rarely have the ability to take meaningful action to improve their privacy, due to a lack of options or effective means of exerting control~\cite{Tabassum2019}. When combined, these factors suggest that residents of today's connected homes may be stuck in a negative cycle of dis-empowerment when faced with the challenge of managing their privacy. 

Motivated by these interrelated problems, we present the results of a 6-week deployment of a \emph{technology probe}, a functional prototype ``to find out about the unknown [...] to hopefully return with interesting data''~\cite{hutchinson2003technology}.  The unknown, in our case, was the design space of \emph{privacy-empowering technologies} for the connected home---future privacy tools that might go beyond static privacy options or labels to support dynamic situational awareness of their informational exposure, and to help end-users build up a rich, conceptual understanding of connected privacy and risks over time. We wanted to explore how such tools might provide targeted, contextualised information relating to their particular situations to maximise relevance, and also to provide the means of taking immediate action, thereby helping users cross both their privacy \emph{gulfs of evaluation} (i.e., understanding their current situation), and \emph{execution} (i.e., taking meaningful action)~\cite{norman2013design}. Technology probes have been seen as ideal for such initial investigations as they constitute ``simple, flexible technologies with three goals: the social science goal of collecting in-context information about the use and the users, the engineering goal of testing the technology, and the design goal of inspiring users and researchers to envision future technologies.''~\cite{O'Brien:2006:HHO:1228175.1228226}.

Through our technology probe, \textit{Aretha}, we sought to jointly explore social science questions relating to privacy risks and preference formation, engineering questions relating to the feasibility of deriving understandable models of data disclosure from network traffic flows, and, finally, design questions pertaining to the capability and interaction design spaces of privacy-empowering technologies.

\section{Background and Motivation}
Our investigation was motivated by both theoretical and empirical findings, including work on privacy conceptual frameworks, preferences, behaviours, and perceptions, especially as such topics relate to the connected home. In this section, we provide a brief overview of the research that most influenced our study and the design of the probe.


Theories of privacy, including conceptual frameworks for understanding privacy in the abstract, can also guide the design of privacy-enhancing technologies, such as by characterising what it might mean to achieve privacy. Conceptualisations of privacy as the control of information about oneself, or `informational self-determination'~\cite{Altman1975,Westin1968}, for instance, frame privacy-enhancing or enabling technology as that which enables people to control how data about them is collected and used, and to determine how their informational selves are represented. The theory of contextual integrity~\cite{nissenbaum2009privacy} (CI) can be applied to explain both the ends and means by which such control might be applied. Using the vocabulary of contexts, actors, attributes and transmission principles, achieving privacy according to CI means successfully regulating the transmission of information to various actors so as to maintain the appropriate sharing norms of the contexts in which they operate~\cite{Apthorpe2018, Martin2016}.


The existence of context-specific norms for information disclosure has been borne out by empirical studies showing that end-users' views on disclosive actions (such as data flows to companies) are influenced by both social and physical context(s) from which such information are gleaned, as well as other factors including the kinds of information being disclosed, perceived norms around disclosures, individuals' relationships with those whom their data are shared, and the purposes of disclosure~\cite{barua2013viewing, Bilogrevic2016, Naeini2017, klasnja2009exploring, lee2016understanding, Leon2013, lin2012expectation}.  When violations of such norms are revealed, the consequences are feelings of betrayal and being ``creeped-out'', whilst when such norms are respected, people feel ``in control'' or benefited~\cite{shklovski2014leakiness,ur2012smart}. Repeated violations of expectations, and the inability to tell whether disclosures are occurring, have been associated with feelings of being perpetually listened to or watched, and even feelings of helplessness and resignation, highlighting potential harmful phenomenological effects associated with long-term violations of privacy~\cite{lau2018alexa}.

The relationship between privacy preferences, intentions, and behaviour is complicated. Early research suggested that people may be divisible into a small set of distinct privacy `types' \cite{Westin1968}, but for decades, studies have noted discrepancies between people's stated preferences and their behaviour. Various attempts have been made to explain (and explain away) these differences  \cite{Acquisti2015a,Kokolakis2017,Smith2011}, with some concluding that researching privacy in experimental isolation is unhelpful~\cite{Dourish2006}, and that privacy attitudes are relatively unpredictable across different scenarios~\cite{woodruff2014would}. This suggests a tension between approaches which aim to classify users according to psychological or behavioural measures in order to predict their privacy choices and adjust default settings accordingly, and those which account for and even encourage the discovery of individual differences. Instead, privacy preferences, intentions and behaviours need to be interpreted alongside the competing pressures and incentives that users face in real-world contexts. As a result, privacy-empowering tools for the connected home cannot divorce the elicitation of preferences and provision of controls from the ebbs and flows of daily life. Beyond \emph{individual} privacy, connected devices are embedded in the idiosyncratic social context of the home, where cohabitants with different needs and preferences must share resources and responsibilities ~\cite{Forlizzi2006,Hargreaves2018,Kraemer2019}.
One additional theory as to why end-users' privacy preferences are seldom predictive is that end-users' preferences are not sufficiently supported by an understanding of privacy risks or awareness to justify their views. Indeed, studies have demonstrated that end-users (and even many experts) lack awareness of all the ways their data is being transmitted and used, whether via the Web, through smartphone platforms and apps, or connected devices in the home~\cite{Zheng2018a}. Such findings are not surprising given that user tracking and information disclosures are, themselves, rarely disclosed in meaningful ways to end-users of connected systems~\cite{Tabassum2019}. Indeed, when information disclosures are made apparent in a form that users can understand, individuals have been shown to articulate more specific and actionable privacy preferences~\cite{gerber2018foxit}, as well as views on ethical, economic (business models), and political dimensions of the data economy~\cite{van2018x,vankleek2017better}.


Increased awareness and understanding of privacy risks, alone, may not result in visible protective changes in behaviour for other reasons. Tabassum et al. found that participants cited trust in manufacturers, faith in regulatory action to protect them, and infeasibility or inavailability of alternatives as reasons for not taking further action despite awareness \cite{Tabassum2019}. Greater awareness of the reality of the nature of data collection, in the absence of available mitigations, might also increase a sense of helplessness and resignation~\cite{van2018x}. Such findings suggest that people need to be aware of, and have strong beliefs about, the efficacy of controls in order to be motivated to act, as demonstrated in the context of security behaviours \cite{Lee2008,Tabassum2019,Wash2015}. 

An alternative view to one that focuses on control over the transmission of data is the social practice informed view that instead promotes accountability
~\cite{Crabtree2017a,Dourish2011b}. Such work has proposed that systems ought to provide affordances to support allowing users to hold manufacturers and service providers accountable for their practices; researchers in human data interaction therefore propose ``liability, agency and negotiability'' as design goals to support such accountability ~\cite{Crabtree2016b,jakobi2018evolving,Tolmie2007}.

Towards increasing both legibility and accountability of individual devices, prior work has proposed privacy labels~\cite{kelley2009nutrition} for apps, websites, and smart home IoT devices~\cite{shen2019iot} that indicate the types and quantities of data that are collected, retained, and disclosed. Privacy fact sheets in buyers guides, such as Mozilla's Privacy Not Included\footnote{\url{foundation.mozilla.org/en/privacynotincluded/}} are a realisation of this idea, providing expert-curated summaries of key aspects of data collection and handling of popular connected devices for the home. Recent work has systematically attempted to identify the factors that should go into these labels, examining how such information factors into purchasing decisions~\cite{emami2019exploring}. But these approaches are limited by the fact that they relate to devices as a whole (as opposed to specific uses/behaviours), and that they are static, meaning they do not allow users to understand how data exposure varies with particular uses, or if exposure profiles change due to changes in apps or software updates. Complementary to this body of work is a collection of tools emerging from the security research community designed primarily for expert users; these were used to inform our probe design as we describe next.

\section{Designing the Technology Probe}

The insights outlined above provide a challenging set of social science, engineering, and design considerations relating to privacy-empowering technologies in the connected home. These drove us to formulate the following key design goals for such technologies, and in turn, our technology probe:

\begin{enumerate}[itemsep=0pt]
    \item \emph{DG1 (Legibility)} - To make the privacy-implicated activities of devices legible to their users, in the forms of both real-time and historical records of all information flows and their destinations.
    
    \item \emph{DG2 (Interpretability)} - To help users interpret and form preferences about data disclosure activities by providing them with a rich conceptual understanding in the technology and business models operating behind their connected devices, including disclosure norms, risks, and purposes. 
    
    \item \emph{DG3 (Actionable Choices)} - To provide \emph{in-situ} privacy controls that enable users to directly take meaningful action on information flows to reduce exposure. 
\end{enumerate}




\subsection{Overview of Existing Tools}
\begin{savenotes}
\begin{table}
\begin{adjustwidth}{-1cm}{-1cm}
    \centering
    {\small
    \begin{tabular}{@{} cl*{10}c @{}}
        & & \multicolumn{10}{c}{} \\[2ex]
        & & \rot{IoT Inspector\footnote{\url{iot-inspector.princeton.edu}}} & \rot{HomeSnitch~\cite{oConnor2019homesnitch}} & \rot{Fing\footnote{\url{www.fing.com}}} & \rot{HomeNet Profiler~\cite{dicioccio2013measuring}} & \rot{IoT Sentinel~\cite{miettinen2017iot}} & \rot{Peek-a-Boo~\cite{acar2018peek}} & \rot{PingPong~\cite{trimananda2019pingpong}} & \rot{IoTSense~\cite{bezawada2018iotsense}} & \rot{HoMonit~\cite{zhang2018homonit}} & \rot{X-Ray Refine~\cite{van2018x}} \\
        \cmidrule{2-12}
        & For end-users         & \CIRCLE & \Circle & \CIRCLE & \RIGHTcircle & \Circle & \Circle & \Circle & \Circle & \Circle & \RIGHTcircle \\
        & Actively Maintained   & \CIRCLE & \Circle & \CIRCLE & \Circle & \Circle & \Circle & \Circle & \Circle & \Circle & \Circle \\
        & Open Source           & \CIRCLE & \Circle & \Circle & \Circle & \CIRCLE & \Circle & \Circle & \Circle & \Circle & \CIRCLE \\
        \cmidrule{2-12}
        & Deep Inspection       & \Circle & \Circle & \Circle & \Circle & \Circle & \Circle & \Circle & \CIRCLE & \Circle & \Circle \\
        & Destination Labelling & \CIRCLE & \Circle & \Circle & \Circle & \Circle & \Circle & \Circle & \Circle & \Circle & \CIRCLE \\
        & Classifies Devices    & \CIRCLE & \CIRCLE & \CIRCLE & \CIRCLE & \CIRCLE & \CIRCLE & \CIRCLE & \CIRCLE & \CIRCLE & \Circle \\
        & Infers Behaviour      & \Circle & \CIRCLE & \Circle & \Circle & \Circle & \CIRCLE & \CIRCLE & \Circle & \CIRCLE & \Circle \\
        & Live Visualisation    & \Circle & \Circle & \Circle & \Circle & \Circle & \Circle & \Circle & \Circle & \Circle & \Circle \\
        & Active Controls       & \Circle & \Circle & \Circle & \Circle & \CIRCLE & \Circle & \Circle & \Circle & \Circle & \Circle \\
        & Anomaly Detection     & \Circle & \CIRCLE & \CIRCLE & \Circle & \Circle & \Circle & \Circle & \Circle & \CIRCLE & \Circle \\
        \cmidrule{2-12}
    \end{tabular}
    }
        \end{adjustwidth}
    \caption{Comparison of Popular Home Network Analysers. Black circles indicate feature presence, with half circles denoting projects designed to be run by end users, but primarily used for research.}
    \label{tab:scanner-comparison}
\end{table}
\end{savenotes}

We began the process of designing our probe by surveying an emerging class of tools for monitoring connected devices in the home. Table \ref{tab:scanner-comparison} summarises the systems and features of each.

Two projects were oriented towards privacy and security-conscious end users: Princeton IoT Inspector and Fing. The former was aimed at a general audience, prioritising ease-of-use and a simple feature set including visualisations of device activity, including a display of data destinations associated with advertising or tracking. The latter was primarily designed for securing and troubleshooting networks, providing tools for advanced users. 

Research tools providing smart device analysis capability fell into three categories. Device fingerprinting in IoTSense~\cite{bezawada2018iotsense} and IoTSentinel~\cite{miettinen2017iot} was able to identify devices based on the network traffic they produced. Behaviour classifiers such as HomeSnitch~\cite{oConnor2019homesnitch}, Peek-a-Boo~\cite{acar2018peek}, PingPong~\cite{trimananda2019pingpong}, and HoMonit~\cite{zhang2018homonit} used machine learning to infer the reasons that devices were sending data to particular destinations at particular times.

A third kind of tool, including HomeNet Profiler~\cite{dicioccio2013measuring} took more holistic approaches rather than device-specific analyses. Homenet Profiler collected MAC addresses and statistics about all of the services advertised by devices across 2,400 French homes. Meanwhile, X-Ray Refine~\cite{van2018x} created data sharing models using static code analysis techniques to build high-level visualisations of smartphone users' entire exposure profiles resulting from their app use. 

While the latter two projects come the closest to the subject of this work, they only address DG1 (\textit{Legibility}), stopping short of providing educational material to increase users' understanding, or providing actionable controls \emph{in-situ} for reducing exposure. This indicates an unexplored design space for privacy-empowering tools that provide real-time legibility, support users to interpret their privacy exposure through better understanding of the technology and business models, and provide active controls. This design space was explored more thoroughly through the design of the Aretha technology probe. Source code for the probe software is freely available under an open source license\footnote{\url{https://github.com/OxfordHCC/Aretha}}.

\subsection{DG1: Legibility}
\begin{figure*}
    \centering
    \includegraphics[width=\linewidth]{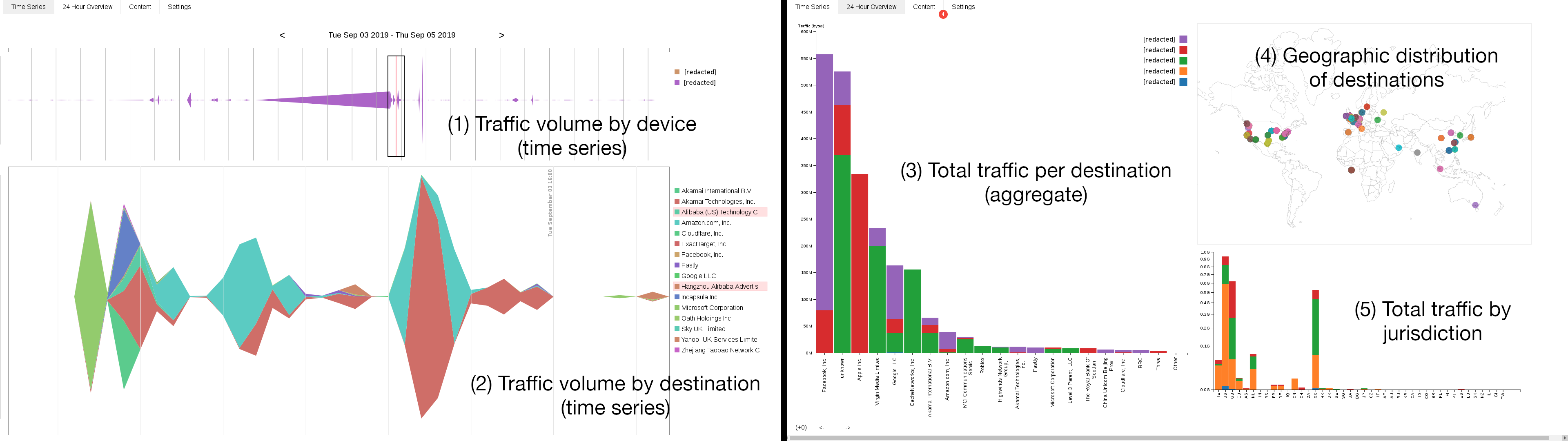}
    \caption{Time series (1-2) and aggregate (3-5) visualisations used in the probe interface.}
    \label{fig:interface}
\end{figure*}


Given the lack of transparency around how connected devices in the home collect and disseminate data, we felt that a crucial first step was to provide them with a view of the ground truth about how their devices disclosed data. Aretha was connected to the home router and acted as a WiFi hotspot, capturing packet headers of traffic passing through it (this reduced the amount of collected data and avoided the challenges of inspecting encrypted payloads~\cite{Ren:2019:IEC:3355369.3355577}). A third party service was used to obtain information about the owner, location, and threat status of external destinations. Unlike IoT Inspector~\cite{huang2019iot}, Aretha included data from other connected devices in the home (e.g. smartphones and tablets) alongside IoT devices.

The interface was based on that of X-ray Refine~\cite{van2018x}, which had been usability-tested in previous lab studies. We adapted this to present information captured from the network both on a timeline, showing the ebb and flow of data throughout the day, and as an aggregated overview showing the total information exposure of the household (see Figure 1).



\subsection{DG2: Interpretability}
Previous studies have established that end-users often lacked the depth of understanding of networked privacy necessary to take views on risks and preventative actions~\cite{kang2015my, xu2012value}. Thus, we felt that in order for end-users to be able to interpret and effective evaluate device disclosure behaviours, Aretha should provide supporting educational and informational material of at least two kinds: essential foundational knowledge about networked privacy concepts and risks, and background information about the specific companies handling their data. To address the first, a short educational curriculum, inspired by FoxIT~\cite{gerber2018foxit}, was assembled from content from the BBC, the UK Information Commissioner, and various other sources, with the aim of providing a broad overview of networked privacy, including Internet basics, how and why devices sent data, and information about data breaches, data protection, and kinds of privacy risks. To provide concrete grounding for essential abstract concepts, concrete examples were constructed, when possible, from users' own home disclosure profiles.  Such examples included illustrations of which devices within their homes sent encrypted versus unencrypted data, and for what purposes these data were being disclosed. Three iterations of the curriculum were piloted via online surveys (total 14 UK residents aged 18 or over), which let us evaluate the inclusion and presentation of different types of information.

\subsection{DG3: Actionable Choices}
The third goal was to enable users to exercise control by being able to take privacy remediation actions directly within the Aretha display.  Towards this end, we faced the challenge of choosing or designing an appropriate method of control; we sought a simple, yet effective method that could serve as a basis for more sophisticated approaches. Since the concept of a firewall was likely to already be familiar and was easily explainable, this seemed a best initial fit. Unlike firewalls for use by experts (which often work at the level of IP or MAC addresses) we designed Aretha's firewall to be user friendly by using device and organisation names in directives (e.g. `block all traffic between <Sam's iPhone> and <Facebook, Inc>'). To make such directives easy to specify, a simple drop-down-based graphical interface was created that directly embedded the ability to create directives into Aretha's visualisations.

\subsection{Pilot Field Test}
To identify usability problems and improve the design of our probe prior to our study, we deployed our initial version within a smart home exhibit at a major building research centre in the UK\footnote{The prototype did not contain the firewall tool.}. Seventeen members of the public were invited to spend time with the prototype system in the fully-equipped smart home test bed, connecting their mobile phones and viewing the data destination visualisations on-screen. Feedback from the pilot was essential in optimising screen layouts to prioritise the visualisations participants found most useful and improving reliability by identifying software and hardware bugs.

\section{Study Methodology}
We designed the main study to consist of a six-week technology probe deployment with three families. As the goals of this study were exploratory, we aimed for a longer-term deployment with a smaller, carefully selected set of households. We felt a six week duration would allow sufficient time for participants to acclimate to having the probe in the home, and for use of the probe to integrate with household rhythms and routines. It was also seen as sufficient time for participants to be exposed to and internalise the educational curriculum which was delivered during the second phase of the study, as well as for members of the household (other than the primary participant) to interact with the probe. All materials from the deployment study can be found at \url{https://osf.io/6j8hc}. The home deployment was approved by our University's IRB, and participants received \pounds200 in shopping vouchers for taking part in the study.

\subsection{Participant Recruitment and Selection}
\begin{table}
    \centering
    \begin{tabularx}{\linewidth}{c|c|c|X}
        \toprule
        P\# & Age & Gender Id. & Education \\ \hline
        P1 & 25-34 & M & Postgraduate Degree \\
        P2 & 35-44 & M & A-Level \\
        P3 & 45-54 & F & Bachelor's Degree \\
    \end{tabularx} \\\vspace{1em}
    \begin{tabularx}{\linewidth}{c|X}
        \toprule
        P\# & Self-Reported Connected Devices in the Home \\ \hline
        P1 & \textbf{Laptop}, \textbf{Phone}, \textbf{Tablet}, Watch \\
        P2 & \textbf{Laptop}, \textbf{Phone}, \textit{Alexa}, TV, Camera, IR, Light, Socket \\
        P3 & \textbf{Laptop}, \textbf{Phone}, \textbf{Alexa}, TV \\
        \bottomrule
    \end{tabularx}
    \caption{Demographic information for the study participants (top) and the types of connected devices they reported owning (bottom). Italics show devices users reported intending to connect to the probe and bold those actually connected according to network traces.}
    \label{tab:demo}
\end{table}

Recruitment was done through a two-step process. First, we recruited an initial pool of interested participants using the Call for Participants platform, reaching out to several smart home hobbyist communities online, and through a mailing list of participants from previous smart home research studies. Interested participants were invited to complete an initial survey with the connected devices they owned, who they shared their living space with, and the geographic area where they lived. Second, we drew on this pool of interested candidates to select three households that represented a variety of technical expertise, family structures, and location of physical residence. Written consent was collected from the main participant in each household, who was required to explain the study to cohabitants and obtain oral consent (we provided written material to help with this). Demographic information on participants is given in Table \ref{tab:demo}.

\subsection{Study Phases and Structure}
At the start of the study, the same researcher visited each household for installation of the probe. The hardware of the probe consisted of an Intel NUC running the Aretha software, an HD display, and a standard keyboard and mouse. The NUC was connected to the participant's broadband router via Ethernet, and was configured to act as a WiFi hotspot to which participants were asked to connect their home devices in lieu of their regular WiFi access point. The interface was always on and continuously visible on the provided display, with users able to toggle between the two screens shown in Fig~\ref{fig:interface}.


Before installation of the equipment, participants were interviewed at home for approximately 30 minutes. This semi-structured entry interview was designed to capture participants' relationships with their connected devices, including questions on their conception of what privacy meant to them, what data they thought their devices shared with whom, and the extent to which they felt in control of their data online. While participants were encouraged to connect as many of their devices as possible to the probe, the initial briefing made it clear that this was not obligatory and that they could redact information collected by Aretha at any time. Additionally, some smart devices required an Ethernet connection and thus could not be connected to the probe.

\begin{figure*}
    \centering
    \begin{tikzpicture}[snake=zigzag, line before snake = 5mm, line after snake = 5mm]
        \draw (0,0) -- (2,0);
        \draw[snake] (2,0) -- (4,0);
        \draw (4,0) -- (13.5,0);
        \draw (0 cm,3pt) -- (0 cm,-3pt);
        \draw (1.5 cm,3pt) -- (1.5 cm,-3pt);
    
        \draw (4.5 cm,3pt) -- (4.5 cm,-3pt);
        \draw (7.5 cm,3pt) -- (7.5 cm,-3pt);
        \draw (10.5 cm,3pt) -- (10.5 cm,-3pt);
        \draw (13.5 cm,3pt) -- (13.5 cm,-3pt);

        \draw (0,0) node[below=3pt] {} node[above=3pt] {Call for Participants};
        \draw (1.5,0) node[below=3pt] {Participant Selection} node[above=3pt] {};
        \draw (4.5,0) node[below=3pt,align=center] {Entry Interview\\Start Stage 1} node[above=3pt] {Week 0};
        \draw (7.5,0) node[below=3pt,align=center] {Check-in Interview\\Start Stage 2} node[above=3pt] {Week 2};
        \draw (10.5,0) node[below=3pt,align=center] {Check-in Interview\\Start Stage 3} node[above=3pt] {Week 4};
        \draw (13.5,0) node[below=3pt,align=center] {Exit Interview\\Equipment Collection} node[above=3pt] {Week 6};
    \end{tikzpicture}
    \caption{Structure of the Home Deployment}
    \label{fig:study-stages}
\end{figure*}
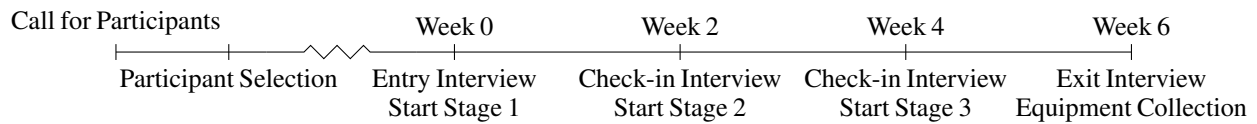

The deployment was split into three stages lasting two weeks each, with participants instructed to interact with the software approximately every other day (see Figure \ref{fig:study-stages}). In the first stage Aretha functioned only as a passive display, encouraging participants to experiment with it to learn more about their devices and to familiarise them with the research software. The second stage added the delivery of the educational curriculum. The final stage of the deployment gave users the ability to use Aretha to block their devices from communicating with companies of their choosing. Participants were contacted at least once during the study to capture feedback and reactions that might have otherwise been forgotten by the end of the six weeks.

Exit interviews were conducted at the conclusion of the study and lasted approximately 40 minutes. Questions included what participants and their cohabitants reactions to the probe were, whether the participants had consciously changed their usage of any devices as a result of the probe, and what they had found most surprising about the data shown by the visualisations. Questions on perception of privacy and control from the entry interview were repeated, with participants given copies of their earlier responses to reflect on and revise if they wished. Participants were given the opportunity to review and redact information recorded by the software, and device MAC addresses were deleted at the guidance of our IRB.

\subsection{Analysis}
Interviews and check-in sessions were transcribed and anonymised by the researcher who conducted the session. Two researchers independently familiarised themselves with and coded the transcripts following \cite{braun2006using}, meeting to review and merge themes to generate the final analysis. A preliminary round of coding was also conducted before the exit interviews to allow emergent themes from the entry and check-in interviews to be further interrogated. Network traces collected by Aretha were referenced during the coding process. While these alone proved to be of limited utility in understanding how the probe was used, they were effective in allowing us to triangulate critical events reported by participants.

\section{Results}
All three probes operated continuously during the 6 week period except that of P1, which experienced a network-card failure that made the system unavailable for approximately four days during the first phase of the study before being rectified by the participant. All probes actively recorded network traffic (packet headers) during the study, recording 112 million packets from 19 devices to 428 unique destinations across 41 jurisdictions worldwide. On average, the top three destinations by volume per participant represented 74\% of the total traffic observed by Aretha. 

Thematic Analysis of the interviews yielded five major themes: \textit{changing conceptions of privacy}---how participants' perceptions of what privacy meant to them changed as the study progressed; \textit{personalisation of privacy concerns}---how participants' understandings of what privacy meant to them became more concretely related to their individual routines and practices; \textit{household members' interest in privacy}---how our participant's cohabitants expressed interest in the probe itself as well as the privacy practices of the household; and \textit{everyday practicalities impacting adoption of the probe}---how participant's usage of the probe (particularly the firewall) was affected by routine happenings in the household. In the following subsections, we briefly outline the journeys taken by each household and describe these themes in the context of each.


\subsection{Household 1}
The first participant lived with his partner in their studio flat, and was technically savvy. P1 was responsible for administering the technology in the household, but online accounts were kept separate between them. The privacy assistant was installed in the main area of their studio apartment, where it was visible throughout the study and used by both the participant and his partner.

Believing that privacy was important to him, P1 had difficulty elaborating further, mentioning tangential practices when asked about the steps he took towards it (such as unsubscribing from email lists and using antivirus software). Going into the study P1 repeatedly voiced concerns that his long term use of the internet meant that he had passed the point of no return with regards to controlling the dissemination of his data online:

\textit{``I've been on the internet for so long that it's just not going to happen. Obviously now I can be more defensive and be cautious, but I just think it's out of control now''}

As new companies appeared on the visualisations, P1 developed his own information gathering process built around the UK national company registrar. Largely unconcerned about the actual data being collected, it was the \textit{relationships} between the companies shown by Aretha that were of importance:

\textit{``So the systematic things I like to do when I research about a company: how many people are involved, what are their sponsors, what's the share price now, how has it changed in the past, what's their next project, what's their previous project, did it change anything, were they successful?''}

During the study P1 changed his broadband supplier, and mentioned being curious about how his choice of ISP might impact his data exposure (network traces showed a 275\% increase in avg. daily traffic to the ISP). Having the probe display in the main area of the apartment allowed it to trigger conversations between P1 and his partner about their data flows. Having not previously been a part of the administration process, his partner was able to educate and inform herself about what their devices were doing:

\textit{``She used to open web pages and then instantly go and check the time series, to just see that Yahoo was open, was there a change?}''

In what became a theme across the participants in the study, P1 found the firewall to be an ineffective way of controlling his data flows. This was mainly due to a fear of it interfering with his digital activities, which were ``everything'', instead leading P1 to seek his own behavioural control mechanisms. The curriculum content about the data economy led him to reflect on the corporations that were indirectly benefiting from his online activities, prompting him to use the ethical track records, details of ownership, and shared interests uncovered during his research to help align his purchasing choices with his responsibilities around sustainability and ethical consumerism:

\textit{``So [the privacy assistant] is doing that by default, and all I'm needing to do from the device is actually to spend my money where everything's fair, everything's fine [...] so I don't want [...] one day delivery from Amazon because they're not paying their people fair. [...] and if I know that when I go on Amazon it's also sending data to X Y and Z, I'm also not going to buy from X Y and Z because that's my grudge against the company now''}

In addition to considering the management of his data that was already online as a problem that was too big to handle, P1 expressed scepticism of GDPR-based data rights as an effective method of controlling his exposure---if behavioural models of him still existed just without his name, then what was the difference? P1's Aretha observed 218 distinct companies operating in 37 countries, highlighting the sheer amount of effort required to meaningfully invoke his data rights. He proposed outsourcing the task to an independent organisation that would be able to handle revocation requests en masse:

``\textit{If someone [...] comes and tells me that we're happy to remove everything from the internet that belongs to you and give you a new birth on the internet [...] then I might be interested to start from scratch}''

\subsection{Household 2}
The second participant lived with his partner and five children in a fully equipped smart home. Many aspects of the house were automated, and P2 was entirely responsible for keeping everything working. The privacy assistant was installed in the living room of the house, allowing for passive observation whilst watching TV or carrying out other leisure activities.

The setup of P2's home was by far the most advanced of the three households, with smart devices touching most aspects of family life. This way of living, to which smart devices were integral, not only foregrounded the reliance of the household on connected technology, but also highlighted how much friction was exerted by adding privacy protecting measures to an already complicated setup. P2 partly envisaged Aretha as a way to validate his own previous privacy and security measures on their home network.

P2 began the study with a more developed conception of privacy, describing it as a matter of `give and take', with him combining the manufacturer's description of what devices were doing with ``a little bit of faith'' that they would operate as specified. While P2 did keep track of the data destinations that appeared over the course of the study, his son was also monitoring the display and holding devices (and P2) to account: 

``\textit{My son is quite technical and is looking at the main bar graph a lot trying to see which bits of traffic are interesting to him, like just now he asked why we're sending so much data to the weather service, which is a good question!}''

Most surprising was both the amount of data sent to large technology firms, outside of the EU (54\% by volume), and the frequency with which data was sent during periods of no or low active device usage. As with the P1, P2 found the delayed impact of blocking decisions interfered with the routines and challenges presented by family life:

``\textit{Because you then have to go away and see what changes happen a little bit later on [...] we've got the five kids running around and sometimes I was reticent to click something in case it affected something and I wouldn't get a chance to come back to it}''

This fear came true when blocking access to a specific company removed comments on YouTube videos. He also noted how the sheer number of destinations observed by Aretha made the firewall difficult to use: \textit{``I found that I got a bit lost on occasion''}. But unlike P1's behavioural control mechanism, P2 did not go on to develop their own control mechanisms in place of the firewall. Instead, P2 mentioned having previously investigated a variety of different options and expressed a lack of control over the family's data flows that did not change over the course of the study.

However, by the end of the study, P2 was able to extend his original definition of what privacy meant to him. This centred around the \textit{purpose} of products and behaviours, particularly how companies balanced meeting user needs against fulfilling other business purposes such as generating advertising revenue, a balance he was adamant needed to remain in the favour of the user. Vocabulary centred around proportionality was used to compare the data flows shown by Aretha with his previous expectations, reflecting that to some extent this was a symptom of the highly interconnected nature of the modern world wide web.

In contrast to P1, P2 drew a clear distinction between information sent to the authorities versus information collected by private companies, and highlighted the importance of keeping insights from devices anonymous and aggregated rather than hyper-personalised: 

``\textit{It's not so much that they know that three thousand people have watched this particular YouTube video, it's do they know that I have watched this YouTube video?}''

\subsection{Household 3}
The third participant lived with her partner and daughter. While the online accounts associated with devices in the house were registered to P3, their partner was responsible for administering devices. The privacy assistant was installed in the hallway between the kitchen and living room of their house, making it easy to glance at throughout the day. 

Similarly to P1, P3 had difficulty describing what privacy meant to her beyond it being important, focusing on a previous bad experience where an organisation had published her personal details without consent. Overall, P3 reported being less engaged with the privacy assistant over the course of the study, often observing new names but being less proactive in investigating them. She found that further research into companies listed as destinations in Aretha was less effective at addressing her concerns. Unlike P1 and P2, P3 was not primarily responsible for the technology in the home, instead deferring to her partner who worked in the IT sector sector, and usage of the privacy assistant between them also appeared to fall into this pattern.

Of surprise were the varied jurisdictions that data was flowing to, and the relationship between action in the house and what was detected by Aretha. This was particularly the case with brand ownership, where activities would show on the visualisation as unexpected companies due to previously unknown corporate structures.

Like the other participants, P3 expressed reluctance to use the firewall due to the risk of unexpected consequences. Unsatisfactory experiences with parental controls when her daughter was growing up factored in to her decision to not use the firewall. The experience did, however, make her much more aware of control mechanisms that she had subconsciously been ignoring. Suddenly she began to notice the cookie opt-out dialogues presented to her on websites, as well as the permission prompts on her phone:

\textit{``So I guess under normal circumstances I wouldn't have cared, but now I'm thinking I don't want you to have access to my phone, I don't want you to have access to my contacts''}

Much more important for P3 though, were the practical steps they could take as a family. Shared discussions about the way the family used and shared online accounts led to a moratorium on account sharing by their daughter, as well proposing the creation of a `bogus' email account that could be used to sign up for apps and services. As a result of these steps P3 reported feeling more in control of their data than at the beginning of the study.

P3's conception of what privacy meant to her also evolved over the course of the study. With some prompts for clarification, she described a purpose-based model similar to that of P2, where data that was extracted from the household needed to be related to a specific purpose that benefited them (such as processing a transaction). This included discussions around data inference, where P3 had used the educational content from the curriculum to develop her understanding of the ecosystem their apps and devices existed in.

\section{Discussion}
\subsection{Revisiting the Design Goals - What Did We Learn?}
Our technology probe aimed to empower people by providing the three kinds of support outlined by our probe design goals. In this section, we revisit these design goals in order to reflect on participants' experiences as they relate to each kind of support, before examining how they interrelate.

With respect to \emph{DG1 (Legibility)}, there was much evidence to support the view that, by the end of the study, households were significantly more aware of data disclosures of their connected devices. In general, each participant had their own expectations of what the probe would show, and subsequently went through a process of reconciling this with the reality shown by the visualisations. While to begin with, participants were mainly interested in \textit{who} devices were sending data to, over the course of the deployment this shifted to \textit{what} information devices were sending and \textit{why} as their awareness and curiosity grew.

Notably, participants tended to focus on the breadth of their data exposure rather than the volume of data going to any particular destination. To an extent this is to be expected given that novel destinations invoke more curiosity than larger traffic magnitudes, but P2 was the only participant that mentioned the `disproportionate' amount of data flowing to Google and Facebook. For the other households, the unfamiliarity of new companies represented a bigger threat than the inference risks of larger ones, even though on average the top three companies for each household accounted for 74\% of the traffic collected by Aretha. Participants were more familiar with these companies, and trust certainly plays into the assessment of risk, but one might have expected more interest in this asymmetry given that user generated privacy strategies in prior work have included depth versus breadth of exposure as a central theme~\cite{van2018x}.

Participant experiences with respect to \emph{DG2 (Interpretability)} were more varied. On one hand, there was evidence to support the view that the educational material supported interpretation and understanding of the network data flows. All participants said that they believed the content in the curriculum had made them think differently about their exposure. The participants who reported engaging more with the curriculum (P1 and P3) also began to incorporate the curriculum concepts and terms into their discussions about privacy, most notably around data sharing. On the other hand, P3 delegated some of the curriculum mini-lessons to her partner, who was the primary individual responsible for managing the technology in the home. The material was also not seen as useful by all participants; P2 felt that he was already aware of the concepts presented and as a result engaged less with the curriculum content.

With respect to \textit{DG3 (Actionable Controls)}, it was clear that the firewall control did \emph{not} meet the intended design goal, but still yielded valuable insights into how future tools might better achieve this. Participants created very few firewall rules to block traffic, even though these controls were made contextually situated, easy to use, and prominent within the assistant.  We believe there were several reasons for this; a primary reason being that there were simply too many data destinations to worry about, a majority of which corresponded to companies unfamiliar to participants. As a result, participants felt they had to evaluate each one individually, and this was simply too much work. The fact that the probe failed to indicate whether it would be `safe' to block destinations also affected firewall usage; when one of P2's blocking actions accidentally disabled essential functionality, they were discouraged from using it further.

Where participants chose their own alternative methods of control, we saw how they took the situational awareness and conceptual grounding that they had been given through Aretha and integrated this with their existing social practices (P1's ethical consumerism and P3's family discussions). Even though P2 did not develop control mechanisms during the study, in the entry interview they expressed a desire to build the probe into their existing security practices to find vulnerabilities in their network. We see that the strength of future tools likely depends on how well they integrate themselves into day to day activities, rather than the raw power they provide as technical tools.

\subsection{Supporting Roles in Privacy Management}
During the study, the probe seemed to serve a variety of different, yet key roles in supporting household privacy management practices. We briefly describe three such roles in this section.

\emph{Formulating and experimenting with privacy preferences and strategies}---A primary role was in supporting the grounding of privacy goals and strategies in terms relating to specific entities, actions and relationships (bringing them closer to the basic ontology of contextual integrity \cite{nissenbaum2009privacy}). At the start of the study, participants expressed concerns in more abstract and general terms, including those corresponding to privacy risks. By the end of the study, however, participants described personalised goals in terms of actors (companies/data controllers), devices, and disclosure actions. Moreover, they were readily able to generate ideas for strategies to achieve their goals, and translate these into actions they could take. Finally, the visualisations facilitated experimentation by providing real-time feedback of the effects of their actions on their exposure, such as accessing certain web sites or using particular apps. Such feedback could be essential for enabling participants to test strategies by assessing the effectiveness of individual actions they might take to achieve their goals (thus supporting the kind of positive feedback loops between beliefs and action noted in prior research \cite{Wash2015}).

\emph{Increasing salience and fostering interest in other householders}---A second important role was the way that the probe brought about not just a heightened awareness of privacy concerns---by being a salient and visible physical totem in their living room---but a transformative one, in which privacy management transitioned from being a solo concern anchored with the individual responsible for technology within the home, to one discussed between multiple home stakeholders. The visibility of the probe within the home inspired conversations among household members, and the visualisations provided a common ground for discussion of disclosures, activities, and devices.

\emph{Supporting accountability of devices, companies, and users themselves}---Participants reported on how legibility (\textit{DG1}), interpretability (\textit{DG2}), and control (\textit{DG3}) did and potentially could further empower them in holding devices, companies, and even each other to account. While each household in the study used the affordances of Aretha for different purposes, the lens of accountability allows us to see how these features are reflected in their use of the probe.

P1 felt empowered through insights and knowledge not only to express clear preferences of not using the services of unethical companies but also stated their preference of not having business with any partners of those companies. They identified these partners by proxy of information flows ``between'' companies from the probe's user interface. In doing so, P1 used the probe to hold \textit{themselves} accountable to their own moral standards, as well as the companies they deemed to be operating unethically (see the Amazon example in the results).


For P2, the probe presented a means by which they intended to hold their \textit{devices} accountable as a way to search for security vulnerabilities in their network. However, they were discontent with the amounts of data being shared with any of the large internet corporations, asking what data was being shared and why. They further expressed the desire to better control these flows of data and hold these companies to account in the future.

The case was different for P3 who was less familiar with the intricacies of connected devices, their data collection, and manufacturers' processing practices; they were not responsible for technology in the family. However, they did engage their partner ``who knew all of this already'' in a conversation on the educational content, and talked to their daughter about sharing accounts with friends.

These examples of reported preferences and behaviour from experiences with Aretha resemble notions of ordinary privacy practices reported in prior research, hinting at the potential for privacy-empowering technologies to ``[enable] people to manage their relationships within the home and with others beyond it''~\cite{Crabtree2017a}. 

\subsection{Informing the Design of Future Home Privacy Tools}
Reflections on the design goals for privacy-empowering tools in the home show the interrelatedness of situational awareness and understanding; information from visualisations and other mechanisms must be tightly interwoven with educational content. A good example of this from the study is the contextual examples provided alongside curriculum modules, providing concrete, personalised examples of how activities in the home related to the more abstract concepts being taught. The shift in questioning from \textit{who} to \textit{why} and \textit{what} presents an opportunity to explore this further. Visualisations and controls based around behaviours rather than destinations place a focus on the types of information that different companies have, and invite questions over what could be done with the information that were missing from the study responses. The design of the `Polly' smart kettle might serve as inspiration for how to combine these~\cite{lindley2017internet}.

The ways that participants used the probe to hold people and companies accountable at a number of different levels suggests that it might be beneficial from a research perspective to highlight this when designing controls. The firewall in the study was concerned with \textit{companies} and \textit{devices}, but controls could similarly be concerned with \textit{people}, \textit{market sectors}, or even entire \textit{corporate structures} (e.g. block data to companies owned by Facebook).

P2's response to the curriculum suggests that future tools could also benefit from tailoring content to the prior knowledge of participants in order to maintain engagement. Additionally, when P3 delegated parts of the learning experience, it exposed the aims of the probe as running counter to the role of P3's partner as the household device administrator---explicitly tailoring a curriculum to each member of the household may offer a means of furthering DG2 in future studies.

\subsubsection{Rethinking Firewalling}
Addressing the challenges identified with DG3, several approaches could make firewalling actions easier and safer. One idea came directly from P1, who suggested outsourcing some of the burden to assistants or experts. One example might be to allow experts to pre-curate custom blacklists that could be selectively enabled and later be overridden by users if necessary. This `ad blocker for the home' approach, which is one taken by tools such as PiHole\footnote{\url{pi-hole.net}}, seems promising and almost obvious as a useful feature in retrospect, and should be strongly considered in future versions of Aretha.

A slightly more sophisticated approach could be providing more powerful operators for allowing sets of destinations to be blocked en masse based on their characteristics (or those of the data flows themselves). Based on the study, useful characteristics would likely include company reputation, jurisdiction, purpose of data collection, retention and disclosure properties. A `learning by example' approach could interactively help people block hosts based on a single action (e.g. ``if you blocked destination X, you might want to block similar destination Y'').

Beyond making it easier, it was clear that what was needed was better visibility and feedback to blocking actions---i.e. a way to make the potential effects and consequences of blocking actions clearer in advance. This stemmed from the ability of the firewall to break the functionality of apps and devices in ways that might not be immediately apparent. The lack of immediate cause and effect increased the difficulty of diagnosing problems, and was further complicated by the fact that only one household member typically had the expertise required to fix problems related to the firewall. The `management by exception' approach in which privacy and security decisions are revisited primarily when problems arise~\cite{jakobi2018evolving} suggests an opportunity for giving people better tools for not only debugging but revisiting and refining their blocking decisions when things break.

\subsubsection{Enhancing Legibility and Accountability}
A key limitation we identified going into the study was the inability for the probe to inspect and make visible the \textit{contents} of traffic flows. Our findings suggest that such a capability would nonetheless be beneficial to improving legibility and accountability. As described above, participants were interested not just in the identities of data controllers, their business models, and reputations, but the contents and sensitivity of what was disclosed, shown by questions pertaining to why the data were captured, and how such data would be later used.  

As the need to harden systems against attackers motivates developers to eliminate the few remaining methods used by researchers to directly intercept encrypted traffic contents, it will be increasingly challenging to reliably realise such an inspection capability. Machine learning driven approaches like HomeSnitch~\cite{oConnor2019homesnitch} that classify device behaviours based on network traffic are promising, but come with several limitations, including that models need to be trained on specific devices and behaviours beforehand, and may not identify the complete contents of such flows.

Enabling end-users to gain an awareness of how data are used, retained, and disclosed by data controllers is another challenge entirely.  We intend to explore the role that legal approaches, including rights given through the EU General Data Protection Regulation (GDPR) might be applied to crowdsource such information in future work.

\subsubsection{Balancing Destinations and Accumulative Risks}
Another limitation of the Aretha probe may have been in the ways it made salient the numbers and identities of destinations, potentially shifting participants' focus away from the large quantities of data being collected by platforms such as Facebook, Amazon, and Google. Although these platforms appeared prominently in every visualisation, they were often not the focus of concern. Future privacy-empowering tools should try to seek a better balance between emphasising destinations and the accumulative privacy risks of data harvested by large platforms.

\section{Limitations and future work}
In line with prior HCI technology probe work (e.g.~\cite{Brown2007,hutchinson2003technology}, we opted for the depth of longer-term case studies with a carefully selected small sample of families. Inherent to the nature of the method, this means that the results can provide holistic understanding of individual appropriation of the technologies (and a resulting design inspiration). However, such a sample is clearly not meant to be representative of the wider population: the observed experiences were likely influenced by selecting early adopters, who were already interested in the aims of the study. It will be interesting to contrast the temporal trajectories and impacts Aretha (and similar technologies) might have on less interested or technology savvy audiences. In addition, future deployments of similar probe-like technologies could be a useful tool with which to unpack the relationships between privacy preference formation (especially across a range of skill and interests), and the shared use of tools by participants. Finally, the particular instantiation of the proposed design goals within Aretha is far from being the only `solution'; nor is it the best solution possible. In the spirit of the technology probe approach, we suggest that our specific design choices are seen only as an early exemplar for a broader class of privacy-empowering tools, hopefully inspiring a variety of different socio-technical ensembles to address the underlying research questions in future. 

\section{Conclusion}

Given their unique contexts, members, and skills, each of our three households went on a different journey during the study. Aretha enabled us to probe the ways they could understand, interact with, and use their connected devices and explore the potential for future privacy-empowering technologies in the connected home. Not all elements of the design goals were successful for all participants, pointing to a range of unmet needs (especially in relation to active controls). The lack of engagement with the firewall was instructive in its own way; while most participants found it difficult to use effectively, due to having already observed, interpreted, and understood the underlying behaviour of their devices they appeared better able to adapt, invent, or imagine other protective mechanisms, tools, and strategies.


By embodying an alternative vision of decentralised, user-empowering IoT infrastructure, Aretha is also a timely reminder that there are alternative ways to engineer the connected home. While some device manufacturers (such as Apple) appear to be embracing more privacy-preserving approaches, end-users may still want an independent means by which they can monitor, audit, and control their devices' disclosures, not only to validate claims made by manufacturers but also to help them better understand how their devices connect to external organisations and systems.

The enormity of the opaque data ecosystem that connected homes are unwittingly tied to could easily be overwhelming. But the results of our probe suggest that all may not be lost; when provided with the right kind of scaffolding, perhaps people can begin to get a handle on the phenomenon, formulate their own informed attitudes and strategies, and better situate themselves within the socio-technical, economic and political realities of the increasingly-connected private realm.

\section{Acknowledgments}
This work was funded by EPSRC grant N02334X/1 and the 2018-2019 ICO Grants Programme.

\balance{}

\bibliographystyle{SIGCHI-Reference-Format}
\bibliography{main}

\end{document}